\documentclass[letterpaper,titlepage,11pt]{article}

\pdfoutput=1

\usepackage{amssymb,amsmath,amsfonts, graphics}
\usepackage{epsfig}
\usepackage{ccaption}
\usepackage{graphicx}

\usepackage[utf8]{inputenc}

\usepackage[
      colorlinks=true,
      linkcolor=blue,
      urlcolor=blue,
      filecolor=black,
      citecolor=red,
      pdfstartview=FitV,
      pdftitle={},
        pdfauthor={Roberto Emparan, Marija Toma\v{s}evi\'c},
        pdfsubject={},
        pdfkeywords={},
        pdfpagemode=None,
        bookmarksopen=true
      ]{hyperref}

\def\clock{{\count0=\time
           \divide\count0 60
           \ifnum\count0<10 0\fi\the\count0
           \multiply\count0 -60 \advance\count0 \time
           :\ifnum\count0<10 0\fi \the\count0
         }}
\newcommand{\timestamp}{{\small\vbox{\hbox{\tt\jobname.tex}
\hbox{\the\day/\the\month/\the\year, \clock}}}}

\newcommand{\beq}{\begin{equation}}
\newcommand{\eeq}{\end{equation}}
\newcommand{\beqa}{\begin{eqnarray}}
\newcommand{\eeqa}{\end{eqnarray}}
\newcommand{\bea}{\begin{eqnarray}}
\newcommand{\eea}{\end{eqnarray}}

\newcommand{\ie}{{\it i.e.,\,}}

\newcommand{\lp}{\left(}
\newcommand{\rp}{\right)}

\newcommand{\Z}{\mathbb{Z}}

\def\v2#1{{\color{blue}{#1}}}

\numberwithin{equation}{section}
\begin{document}

\begin{titlepage}
\leftline{}%{\timestamp}
\vskip 2cm
\centerline{\LARGE \bf Strong cosmic censorship in the BTZ black hole}
\vskip 1.6cm
\centerline{\bf Roberto Emparan$^{a,b}$, Marija Toma$\check{\textrm{s}}$evi\'c$^{a,c}$}
\vskip 0.5cm
\centerline{\sl $^{a}$Departament de F{\'\i}sica Qu\`antica i Astrof\'{\i}sica, Institut de
Ci\`encies del Cosmos,}
\centerline{\sl  Universitat de
Barcelona, Mart\'{\i} i Franqu\`es 1, E-08028 Barcelona, Spain}
\smallskip
\centerline{\sl $^{b}$Instituci\'o Catalana de Recerca i Estudis
Avan\c cats (ICREA)}
\centerline{\sl Passeig Llu\'{\i}s Companys 23, E-08010 Barcelona, Spain}
\smallskip
\centerline{\sl $^{c}$Kavli Institute for Theoretical Physics}
\centerline{\sl University of California, Santa Barbara, CA 93106}
\vskip 0.5cm
\centerline{\small\tt emparan@ub.edu,\, mtomasevic@icc.ub.edu} 

\vskip 1.6cm
\centerline{\bf Abstract} \vskip 0.2cm \noindent
Recently it has been shown that quantum fields can be regular on the inner, Cauchy horizon of a rotating BTZ black hole, which appears to indicate a failure of strong cosmic censorship. We argue that, instead, what these results imply is that the inner horizon remains non-singular when leading-order backreaction of the quantum fields is computed, but, after next-order backreaction is accounted for, it will become singular. Then, strong cosmic censorship will be enforced in the BTZ black hole. We support our claims using a four-dimensional holographic dual of the system, which connects the instability of the inner horizon of the BTZ black hole to that of Kerr-type black holes.

\end{titlepage}
\pagestyle{empty}
\small
%\tableofcontents
\normalsize
\newpage
\pagestyle{plain}
\setcounter{page}{1}

\section{Introduction}

Solutions for charged or rotating black holes typically contain, in addition to the outer event horizon, an inner horizon that gives rise to perplexing features. Seemingly, particles from the exterior of the black hole could be sent across the inner horizon to emerge in another exterior universe; if an observer could cross this horizon, they would see a timelike singularity in their past; and, as the crossing moment is approached, they would witness the entire history of the exterior universe compressed in a finite duration of their proper time \cite{Simpson:1973ua}.

These phenomena are troublesome for physics. In Anti-deSitter space, the transmission of signals from one asymptotic region to another would be in conflict with holographic quantum gravity \cite{Engelhardt:2015gla}; the singularity in the past is a sign of the lack of predictivity of the classical theory beyond the inner horizon, which is indeed a Cauchy horizon; and the infinitely blue-shifted energy of any excitation coming from the exterior should generate a strong backreaction at the inner horizon \cite{Simpson:1973ua,Poisson:1990eh}. The latter effect suggests the notion of strong cosmic censorship \cite{Penrose:1974,Penrose:1979}: the smooth Cauchy horizon is an idealized artifact of the exact solutions, and any generic perturbation (classical, but also possibly quantum) must destabilize it and turn it into an impassable barrier that disposes of the troubles that would otherwise follow.

The question of whether such censorship actually holds has been revisited recently from different angles (reviewed in detail in \cite{Hollands:2019whz}). For our purposes here, the studies in \cite{Lanir:2018vgb,Zilberman:2019buh,Dias:2019ery,Papadodimas:2019msp,Balasubramanian:2019qwk,Hollands:2019whz} are particularly relevant. The latest of these \cite{Hollands:2019whz} finds that the expectation value of the quantum stress-energy tensor for a free scalar field generically diverges as the inner horizon is approached from the outside, behaving as
\beq\label{divT}
\langle T_{VV}\rangle \sim \frac{C}{V^2}
\eeq
with the null coordinate $V\to 0^-$ at the Cauchy horizon, and $C$ a constant that depends on black hole parameters. This divergence is equal or stronger than that from classical field effects. Although the analysis is done explicitly for the class of Reissner-Nordstr\"{o}m solutions with a cosmological constant, according to \cite{Hollands:2019whz} it can be adapted to yield the same conclusions when rotation is present. Other quantum probes of the inner horizon reach essentially the same results \cite{Papadodimas:2019msp,Balasubramanian:2019qwk}. This lends great confidence to the idea that strong cosmic censorship is enforced by quantum physics when classical physics may not do it.

However, these works also concur on an exception to this conclusion, first identified in \cite{Dias:2019ery}: the inner horizon of the rotating BTZ black hole in three dimensions \cite{Banados:1992wn,Banados:1992gq}. The highly symmetric nature of this black hole, which is a discrete quotient of AdS$_3$, imposes precise cancellations in the coefficient $C$ of the divergence of the stress tensor, rendering it mild enough that its backreaction should leave the inner horizon smooth. Even though the BTZ black hole does not have any curvature singularity past the inner horizon, the difficulties discussed above should still be present. The potential conflict with \cite{Engelhardt:2015gla} would seem worrisome for the consistency of a theory of quantum gravity. More generally, it would be unclear what the correct physics is beyond the inner horizon \cite{Dias:2019ery}.

\section{Quantum backreaction}

Let us reexamine what the results of these works imply. Backreaction from the quantum fields is accounted for by solving the Einstein equations for a perturbation of the BTZ black hole sourced by the renormalized stress-energy tensor.  If this were divergent at the inner horizon, then the correction to the geometry would become large. Although perturbation theory would cease to be valid in that region, it is clear that the new geometry would be significantly changed, likely creating high curvatures outside the inner horizon. If, instead, the stress tensor is finite, as \cite{Dias:2019ery,Hollands:2019whz} find for the BTZ black hole, then the backreaction at the inner horizon will be moderate, with no signs of a singularity. In order to fully account for the backreaction one would need to solve simultaneously both the Einstein equations and the quantum field theory as a coupled system, but this is too hard a problem. Nevertheless, one may take the perturbative study to the next order by solving for the quantum field theory in the first-order-corrected black hole geometry, and then backreact again with the resulting renormalized stress tensor.

It is at this stage that the situation will change at the inner horizon of the BTZ black hole. The quantum-corrected geometry where we must recalculate the renormalized quantum stress tensor will not possess the same high degree of symmetry of the initial classical solution. In particular, it will no longer be a spacetime of constant negative curvature with discrete identifications. In the absence of any symmetry protection, the stress tensor at the inner horizon is expected to diverge in the generic form \eqref{divT} with a non-zero coefficient $C$, since the quantum corrections will affect it. This will then create a large backreaction on the inner horizon, thus implementing strong cosmic censorship.

The first-order perturbative backreaction effects of the quantum stress tensor of a free conformal scalar in BTZ  \cite{Steif:1993zv} have not been properly derived yet\footnote{The calculations in \cite{Casals:2016odj,Casals:2019jfo}, which found a singular inner horizon, have been criticized in \cite{Dias:2019ery} as making improper use of the stress tensor beyond the Cauchy horizon.}. As we have explained, they are expected to leave the inner horizon geometry non-singular, but also altered from its highly symmetric initial form. Here, we will follow a different approach to solving this problem, one which supports our arguments above, while also casting them into a different, helpful light.

\section{Holographic dual analysis}

We employ a holographic approach to solving the quantum field theory in the BTZ black hole background and computing its gravitational backreaction, following the work of \cite{Emparan:1999fd,Emparan:2002px}.

\subsection{The model}

It was argued in \cite{Emparan:2002px} that the three-dimensional system of the BTZ black hole and a conformal quantum field theory in its presence is dual to a four-dimensional solution of the Einstein theory with a negative cosmological constant, specifically the AdS C-metric with rotation. This spacetime contains a black hole that follows an accelerating trajectory in AdS$_4$. In the construction of \cite{Emparan:1999fd}, part of the spacetime is cut off by a brane that slices across the black hole horizon, so the $(2+1)$-dimensional geometry induced on the brane has a black hole. Via holography \cite{Verlinde:1999fy,Gubser:1999vj,Karch:2000ct}, such geometries provide solutions to the Einstein equations
\beq
R_{\mu\nu}-\frac12 R g_{\mu\nu}-\frac{1}{\ell_3^2} g_{\mu\nu}=8\pi G_3 \langle T_{\mu\nu} \rangle\,,
\eeq
where $g_{\mu\nu}$ is the three-dimensional metric induced on the brane, $\ell_3$ and $G_3$ are the effective three-dimensional AdS radius and Newton's constant, and $\langle T_{\mu\nu} \rangle$ is the renormalized stress energy tensor of the quantum conformal fields (with a cutoff) that are dual to the four-dimensional bulk gravity.

When the bulk theory is classical, the results for the quantum CFT correspond to the leading order in a $1/N$ expansion, where $N$ is a measure of the number of strongly-coupled conformal degrees of freedom in the field theory.\footnote{Actually, this would be $\sim N^{3/2}$ for the CFT on the worldvolume of $N$ M2-branes dual to AdS$_4$.} We emphasize that the classical bulk solution yields the complete gravitational backreaction of the CFT to leading order in the $1/N$ expansion. That is, to this order, these effects are included not as linearized gravitational perturbations but fully non-linear effects. This is an improvement over the conventional perturbative approach to backreaction that we discussed above. Note also that the leading order in the $1/N$ expansion consists of planar diagrams with arbitrary number of loops, so it is not the same as the perturbative loop expansion.

There is one subtlety in this dual construction when the theory on the brane has a negative cosmological constant, $\ell_3\neq 0$. Namely, there is no massless graviton localized on the brane, but a massive graviton bound state \cite{Karch:2000ct}. As a result, gravity on the brane behaves in a three-dimensional way up to distance scales of the order of the inverse of the bound state mass, but becomes four-dimensional at longer scales. However, we do not find any indication that this is relevant for our analysis.\footnote{One might suppose that this is why there is an upper bound on the mass and area of the black holes localized on these branes \cite{Emparan:1999fd}. However, the holographic calculations of \cite{Hubeny:2009rc}, discussed below, also find an upper bound on the mass, even though the geometry of the BTZ black hole in that model is fixed and not dynamical. Note also that \cite{Emparan:1999fd} dealt with these infrared effects in another manner, introducing a second brane.}

\subsection{Quantum-corrected BTZ}

The construction of \cite{Emparan:1999fd} allows a very explicit description of the solution and its properties, of which we will only give the details that are needed here. The geometry on the brane, which we refer to as the quantum-corrected BTZ black hole (qc-BTZ), takes the form
\beq\label{qcBTZ}
ds^2=-\lp \frac{r^2}{\ell_3^2}-M+\frac{J^2}{4 r^2}-\frac{\alpha(M,J)}{r}\rp dt^2+\frac{dr^2}{\frac{r^2}{\ell_3^2}-M+\frac{J^2}{4 r^2}-\frac{\alpha(M,J)}{r}}+r^2\lp d\phi-\frac{J}{2r^2}dt\rp^2\,,
\eeq
where we have set $G_3=1/8$. The quantum corrections from the CFT are encoded in $\alpha(M,J)$. The explicit form will be given in \cite{toappear},, but it is not particularly illuminating and in any case we will not require it. Suffice to say that $\alpha>0$.

This spacetime does not satisfy the Einstein-AdS equations, $R_{\mu\nu}=-(2/\ell_3^2)g_{\mu\nu}$, but it is straightforward to identify the stress-energy tensor that sources it,\footnote{Since gravity on the brane is dynamical, we can obtain the holographic stress tensor by simply extracting the `right-hand side' of the Einstein equations on the brane.}
\beq\label{qcBTZstress}
\langle T^\mu_\nu\rangle =\frac{\alpha(M,J)}{2\pi r^3}\textrm{diag}(1,1,-2)+ \frac{3J\alpha(M,J)}{4\pi r^5}\delta^\mu_t\delta^\phi_\nu\,.
\eeq

This result can be compared with two other computations of the quantum stress-energy tensor in the BTZ black hole: \cite{Steif:1993zv} for a free conformal scalar field, and \cite{Hubeny:2009rc} for a holographic CFT dual to a four-dimensional bulk. It is important to note that the use of the AdS/CFT duality in the latter is very different than ours, specifically in two respects: (i) the BTZ geometry in \cite{Hubeny:2009rc} lies at the asymptotic boundary of AdS$_4$, so there is no gravitational backreaction on it; (ii) although the bulk solution employed in \cite{Hubeny:2009rc} is obtained from a rotating black hole, namely the Kerr-AdS$_4$ solution, the construction involves double Wick rotations and bulk coordinate transformations, with effect that the horizons of the boundary BTZ black hole are not part of---\ie slices of---the horizons of the original Kerr-AdS$_4$ black hole. We will return to these points in our later discussion.

Let us first compare these three calculations when $J=0$. Then, the results of \cite{Steif:1993zv} and \cite{Hubeny:2009rc} have the same structure as \eqref{qcBTZstress}, only with different functions $\alpha(M)$. The latter is expected: \cite{Steif:1993zv} refers to different conformal matter, and although in the two holographic setups the CFT is the same, \eqref{qcBTZstress} includes self-consistently its backreaction on the geometry \eqref{qcBTZ}, which \cite{Hubeny:2009rc} does not. Since only $\alpha(M)$ changes between these results, it is natural to conjecture that the stress tensor of any conformal field in the static BTZ black hole will have the same structure, and that the backreacted geometry will generically take the form of \eqref{qcBTZ} with $J=0$.

When $J\neq 0$ the holographic result of \cite{Hubeny:2009rc} has the same form as  \eqref{qcBTZstress}, but the one in \cite{Steif:1993zv} reproduces only some aspects for small $J$. The differences for finite $J$ are naturally attributed to weak vs.~strong coupling effects.

The quantum-corrected properties of the horizons in \eqref{qcBTZ} are easily computed. For instance, their temperatures and angular velocities are
\beqa
T_\pm&=&\pm\frac1{2\pi}\lp \frac{r_\pm}{\ell_3^2} -\frac{J^2}{4r_\pm^3}+\frac{\alpha}{2r_\pm^2}\rp,\\
\Omega_\pm&=&\frac{J}{2 r_\pm^2}\,.
\eeqa
Here $r_\pm$ are the quantum-corrected horizon radii in \eqref{qcBTZ}, which are the positive roots of the quartic equation
\beq
\frac{r^4}{\ell_3^2}-Mr^2+\frac{J^2}{4}-\alpha r=0\,.
\eeq
If $\alpha$ is small, then
\beq
r_\pm =r_\pm^0\pm\frac{\alpha}{2\sqrt{M^2-J^2/\ell_3^2}}+O(\alpha^2)\,,
\eeq
where the classical BTZ values are
\beq
r_\pm^0=\ell_3\lp\frac{M\pm\sqrt{M^2-J^2/\ell_3^2}}2\rp^{1/2}\,.
\eeq
It is easy to see that the quantum corrections raise the temperature of the horizons for all $\alpha>0$, which is in line with the observed reduction of their entropy \cite{Emparan:1999fd}.

The qc-BTZ solution \eqref{qcBTZ} has a curvature singularity at $r=0$, which is a ring singularity when $J\neq 0$.\footnote{This is clearer after changing $t\to t-J\phi/2$, see \cite{Emparan:1999fd}.} But the most important feature for us is that the geometry \eqref{qcBTZ} and the stress tensor \eqref{qcBTZstress} are smooth at the inner horizon $r=r_-$. So we may be led to conclude that strong cosmic censorship is violated in this black hole.

However, as we discussed, this construction only yields the quantum-corrected geometry to leading-order in the $1/N$ expansion of the conformal theory (planar diagrams), for which the dual gravitational bulk theory is purely classical. At finite order in $N$, the effects of quantum fields in the bulk---at the very least, graviton loops---must be included. Such effects are not easy to compute, but we will not need them explicitly in order to reach our main conclusion.

\subsection{Strong cosmic censorship}

Let us consider the properties of the four-dimensional bulk solution which on a brane slice yields the qc-BTZ geometry \eqref{qcBTZ}. It was shown in \cite{Emparan:1999fd} that this four-dimensional black hole has a structure qualitatively like that of the four-dimensional Kerr black hole (or Kerr-AdS$_4$): a ring singularity, and regular inner and outer horizons, which in the brane section \eqref{qcBTZ} are at $r=0$ and $r=r_\pm$. Indeed, the Kerr black hole is recovered in the limit that the bulk black hole is small, while for larger size it has a pancaked shape; and the Kerr-AdS$_4$ solution is recovered when the brane tension is sent to zero.

What will the effect be of quantum fields that propagate in this four-dimensional geometry? Given the similarity of the bulk black hole with the Kerr solution, we expect the generic divergent behavior \eqref{divT} at the inner horizon. As a consequence, the backreaction of this bulk quantum effect will be enough to make the Cauchy horizon singular, in the bulk and also on the brane section, therefore implementing strong cosmic censorship in the BTZ black hole. 

An apparent difficulty for extending the methods and conclusions of \cite{Hollands:2019whz} to our setup refers to the boundary conditions for the field, which differ from those that one would impose in Kerr of Kerr-AdS$_4$. In the presence of the brane, the natural boundary condition for the bulk fields is $\Z_2$-orbifold symmetry. Admittedly, a specific analysis would be needed to establish conclusively that this modification does not entail a cancellation of the divergence \eqref{divT}, but we find this very unlikely. The study of \cite{Hollands:2019whz} strongly suggests that this behavior is generic except in very special, highly symmetric situations that fine-tune it away. The brane construction in the rotating AdS C-metric is, if anything, less symmetric than the Kerr solutions: unlike the latter, it does not have a Killing tensor; the $\Z_2$ symmetry on the brane is also present in Kerr-AdS$_4$ as reflection symmetry on the equatorial plane---this plane is actually the `tensionless brane' limit of the brane construction; and, in addition, the asymptotic boundary in the AdS C-metric is a distorted version (not conformally flat) of the one in Kerr-AdS$_4$. So, since all the boundary conditions in our bulk geometry are similar but less symmetric than in the Kerr solutions, there do not appear to be the conditions for special protection against any divergences.

Observe that we may also invoke the (extended) conclusion of \cite{Hollands:2019whz} directly for the three-dimensional geometry \eqref{qcBTZ}, and possibly similarly quantum-corrected BTZ black holes even when they do not have a holographic bulk counterpart. Our dual construction strengthens the argument by placing it in the class of instabilities of inner horizons of Kerr-type black holes. In this regard, we have invoked bulk quantum effects since they yield large and robust divergences. But classical gravitational perturbations in the Kerr solutions are also expected to become singular on the Cauchy horizon, even though their strength and the extent to which they enforce strong cosmic censorship is a subtle matter \cite{Dafermos:2017dbw}. It is very plausible that classical perturbations in our bulk geometry will develop a similar singularity which would extend to the brane black hole. From the three-dimensional point of view, these would be perturbations of the quantum CFT state at leading (planar) order in $1/N$. That is, the inner horizon of the qc-BTZ black hole could be unstable to quantum fluctuations already at leading order in $1/N$. 

Finally, it is interesting to discuss an apparently similar holographic reasoning put forward in \cite{Dias:2019ery}. There it was observed that the holographic CFT stress tensor in the BTZ black hole obtained in \cite{Hubeny:2009rc} (which, as we discussed above, has the same form as \eqref{qcBTZstress}) is finite at the black hole inner horizon. It was then speculated that $1/N$ corrections---\ie quantum bulk effects---in that model could spoil the smoothness of the Cauchy horizon. If this were the case, the implementation of strong cosmic censorship would not require including backreaction on the geometry. This is certainly an interesting possibility, already suggested in \cite{Dias:2019ery,Hollands:2019whz}, but it is not clear how to argue for it in the model of \cite{Hubeny:2009rc}. While it may be interesting to investigate this further, we believe that it is not necessary to do so in order to conclude in favor of strong cosmic censorship in BTZ, once the consequences of backreaction are factored in.

\section{Conclusion}

Our argument is easily summarized: the cancellations that protect the finiteness of the renormalized stress tensor at the BTZ inner horizon are very fragile, and cannot be expected to survive when backreaction effects are included beyond the leading order. The generic divergence found in \cite{Hollands:2019whz} is then expected to prevail. A holographic dual construction gives us a concrete result for how the first-order backreaction changes the geometry, and also specific expectations for higher orders: although the inner horizon of the qc-BTZ solution is indeed smooth, through its bulk dual it will be as sensitive to quantum instability as it is in the Kerr black hole.

So we conclude that the BTZ black hole is not an outlier: strong cosmic censorship must apply to it as much as it does to other higher-dimensional black holes.

%\addtocontents{toc}{\protect\setcounter{tocdepth}{2}}

%\tableofcontents
%\newpage

\section*{Acknowledgments}

We are grateful to {\'O}scar Dias, Gary Horowitz, Veronika Hubeny, Don Marolf, Harvey Reall, Jorge Santos, and Benson Way for useful discussions, and to the KITP, UC Santa Barbara, for warm hospitality during the program ``Gravitational Holography''. Work supported by ERC Advanced Grant GravBHs-692951 and MEC grant FPA2016-76005-C2-2-P. This research was supported in part by the National Science Foundation under Grant No. NSF PHY-1748958.

%%%%%%%%%%%%%%%%%%%%%%%%%%%%%%%%%%%%%%%%%

\end{document}